\documentclass[a4paper,11pt]{article}
\usepackage{amsfonts}
\usepackage{amsmath}
\usepackage{latexsym}
\usepackage[dvips]{graphicx}
\begin{document}
\newcommand{\be}{\begin{eqnarray}}
\newcommand{\ee}{\end{eqnarray}}
\newcommand{\al}{\alpha}
\newcommand{\bt}{\beta}
\newcommand{\dt}{\delta}
\newcommand{\ep}{\epsilon}
\newcommand{\om}{\omega}
\newcommand{\p}{\partial}
\newcommand{\ta}{\theta}
\newcommand{\tx}{\tilde{x}}
\begin{flushright}
ITFA-99-28
\end{flushright}
\begin{center}
\Large{Einstein and the Kaluza-Klein particle}\\
\vspace{.25in}
\large{Jeroen van Dongen\footnote{jvdongen@wins.uva.nl; jvdongen@feynman.princeton.edu}\\
\vspace{.35in}
Institute for Theoretical Physics, University of Amsterdam\\
Valckeniersstraat 65 1018 XE Amsterdam\\
and \\
Joseph Henry Laboratories, Princeton University\\
Princeton NJ 08544}  \\
\begin{abstract}

In his search for a unified field theory that could undercut quantum mechanics, Einstein considered five dimensional classical Kaluza-Klein theory. He studied this theory most intensively during the years 1938-1943. One of his primary objectives was finding a non-singular particle solution. In the full theory this search got frustrated and in the $x^5$-independent theory Einstein, together with Pauli, argued it would be impossible to find these structures. 
\end{abstract}
\end{center}

\section{Introduction} 

After having formulated general relativity Albert Einstein did not immediately focus on the unification of electromagnetism and gravity in a classical field theory - the issue that would characterize much of his later work. It was still an open question to him whether relativity and electrodynamics together would cast light on the problem of the structure of matter \cite{ein:15}. 
Rather, in a 1916 paper on gravitational waves he anticipated a different development: since the electron in its atomic orbit would radiate gravitationally, something that cannot \emph{``occur in reality''}, he expected quantum theory would have to change not only the \emph{''Maxwellian electrodynamics, but also the new theory of gravitation''} \cite{ein:16}\footnote{Our translation of the German original.}. Einstein's position, however, gradually changed.
From about 1919 onwards, he took a strong interest in the unification programme\footnote{For an account of the early years of the unified field theory programme, see \cite{viz:94}.}. In later years, after about 1926, he hoped that he would find a particular classical unified field theory that could undercut quantum theory. Such a theory would have to contain the material objects -sources and fields- and their dynamics. He would even expect the distinction between these concepts to fade: \emph{``a complete field theory knows only fields and not the concepts of particle and motion''} \cite{ein:35}. We will study how he wanted to realize these principles in classical Kaluza-Klein theory, and try to see what his objectives and results were.

In 1920, when giving his inaugural lecture in Leyden, Einstein for the first time publicly commented positively on the unification programme:\\
\\
\emph{``It would be a great step forward to unify in a single picture the 
gravitational and electro-magnetic fields. Then there would be a worthy completion 
of the epoch of theoretical physics begun by Faraday and Maxwell...''} \cite{ein:20b}, as in Vizgin \cite{viz:94}.\\
\\
Wolfgang Pauli had at first been a strong proponent of classical unification, in 
particular of Weyl's conformal theory \cite{viz:94}. However, during 1920 his opinion shifted to a more critical position and in a public discussion during the 86th ``Naturvorscherversammlung'' in Nauheim he addresses 
Einstein: \\
\\
 \emph{``I would like to ask Einstein if he agrees that one can only expect a 
solution to the problem of matter by a modification of our ideas about space (and 
perhaps also time) and electric fields in the sense of atomism...or should one 
hold on to the rudiments of the continuum theories?''} \cite{pau:20}\footnotemark[2]\\
\\
Einstein would then not decide either way and replies with a subtle 
epistemological remark, concerning the situation when a scientific concept or 
theory is in opposition with nature\footnote{He makes a distinction between a 
procedure where a concept is dropped from a physical theory, and a procedure where 
the system of arrangement of concepts to events is replaced by a more complicated 
one, that refers to still the same concepts \cite{ein:20}.}.  He would however 
keep working on (several) continuum theories, letting future success decide 
which approach would be justified \cite{ein:21,viz:94}. Pauli clearly formulated his criticism in his encyclopedia article on 
relativity, which also sums up the most important characteristics of the 
unified field theory programme. \\
\\
\emph{``It is the aim of all continuum theories to derive the atomic nature of 
electricity from the property that the differential equations expressing the 
physical laws have only a discrete number of solutions which are everywhere 
regular, static and spherically symmetric....}[however] \emph{the existence of 
atomicity should in theory be interpreted in a simple and elementary manner and should not...appear as a trick in analysis.''} \cite{pau:58}\\
\\
One sees the essential role particle solutions play in the unified field theory programme.
To resolve the problem of matter Pauli felt one needed to abandon the approach characteristic of classical field theories.\\
\\
\emph{``New elements foreign to the continuum must be added to the basic structure 
of the theory before one can arrive at a satisfactory solution of the problem of 
matter.''} \cite{pau:58}\\
\\
Pauli exhibits a clear sense for future developments concerning the quantized nature of matter. Einstein would not give up on classical field theory:\\
\\
\emph{``Before we seriously start considering such far-fetched possibilities we have to find out whether really follows from the recent efforts and facts that it is impossible to succeed with partial differential equations.''} \cite{ein:23} \footnotemark[2]

\section{Kaluza's theory}

In 1921 Einstein presented to the Prussian Academy a paper by Theodor Kaluza, entitled ``Zum Unit\"atsproblem der Physik'', in which the gravitational and electro-magnetic field are geometrically unified in five dimensions. Einstein had known of Kaluza's idea already in 1919 and had commented positively on it in a letter to Kaluza (for correspondence of Einstein to Kaluza, see \cite{des:83}).
Kaluza had proposed a five dimensional theory of relativity, so that one could 
geometrically unify gravity and electromagnetism \cite{kal:21}. 
He started out by taking as 
action the five dimensional Ricci scalar, and he imposed the cylinder condition, 
i.e. the components of the metric $g_{IJ}$ should not depend on the space-like fifth dimension. He then interpreted the $5\times 5$ metric as follows\footnote{Capital Latin indices run over 0,1,2,3,5, Greek indices run over the usual dimensions 0,1,2,3. Later we will use lower-case Latin indices, these run over the space-like directions 1,2,3.}; 
\be  
g^{(5)}_{IJ}= \left( \begin{array}{rr}
g^{(4)}_{\mu \nu} &  \alpha A_{\nu } \\
 \alpha A_{\mu} & 2 \phi
\end{array} \right)   \,\,\,\,\,\,\,\,\,\,\,\, \mbox{with} 
\,\,\,\,\,\,\p_{5}g_{IJ}=0,
\ee
where $g_{\mu\nu}$ represents the metric of the four dimensional spacetime, 
$A_{\mu}$ is the gauge field from electrodynamics and $\alpha$ is a coupling constant, related to the Newton constant $\kappa$ via $\alpha=\sqrt{2 \kappa}$. There is a new field $\phi(x)$, often called the dilaton.
Kaluza evaluated 
the Ricci tensor for linearized fields:
\be 
\begin{array}{ll}
R_{\mu\nu}=\partial_{\lambda} \Gamma^{\lambda}_{\mu \nu},\\
R_{5\nu}=- \alpha \p^{\mu}F_{\mu \nu},  \label{eq:max}\\
R_{55}=-\Box \phi, 
\end{array} 
\ee
and continued by studying the specific example of a charged particle in the five dimensional space. This taught him that the momentum in the fifth direction needs to be interpreted as the electric charge: $j^{\nu}=\kappa T^{\nu 5}$. The charge is a conserved quantity by virtue of the translation symmetry on $x^5$. 
From the five dimensional line element Kaluza recovered for the equation of motion of a charged particle:
\be 
m\Big(\frac{d^2 x^{\mu}}{d\tau^2}+\Gamma^{\mu}_{\beta \gamma}\frac{dx^{\beta}}{d\tau}\frac{dx^{\gamma}}{d\tau}\Big) =e \,F^{\mu}\,_{\nu}\frac{dx^{\nu}}{d\tau}
-\frac{1}{2\kappa}\frac{e^2}{m} \p^{\mu} \phi
\ee
which gives 
the Lorentz force law under the assumption of small specific charge $e/m$. 
However, for the electron this assumption breaks down and contrary to experience, the interaction with the scalar field becomes the leading term in the equation of motion - pointed out to Kaluza by Einstein \cite{kal:21}. To Kaluza this was an important shortcoming of the theory. 

The theory Kaluza 
formulated was, up to the point where he introduced the particle, a vacuum theory. 
It did not contain an additional term in the action besides the five dimensional Ricci scalar: the field equations are given by equating the Ricci tensor (\ref{eq:max}) to zero.
 But once the particle is introduced, and consequently its 
energy-momentum tensor, one has left the vacuum theory. 
So when Kaluza in his example writes $R_{MN}= \kappa T_{MN}$, one wonders where the
source term should come from, if it would not be introduced as an additional 
matter-term in the Lagrangian. In fact, Einstein asks the same question in a 
paper with Jakob Grommer:\\
\\
\emph{``...Kaluza introduces besides the quantities $g_{IJ}$ another tensor 
representing the material current. But it is clear that the introduction of such a 
tensor is only intended to give a provisional, sheer phenomenological description 
of matter, as we presently have in mind as ultimate goal a pure field theory, in 
which the field variables produce the field of `empty space' as well as the 
electrical elementary particles that constitute 'matter'.''} \cite{ein:22} \footnotemark[2]\\
\\
We will see that Einstein would want the sources to come from the geometry. 
Einstein and Grommer find that Kaluza's vacuum field equations ($R^{(5)}_{IJ}=0$)  cannot produce non-singular rotation symmetric particle solutions. 
It seems plausible that because of this argument Einstein 
would, for about five years, no longer work on Kaluza's theory. A strong 
indication is the following letter to Weyl, written in 1922; \\
\\
\emph{``Kaluza seems to me to have come closest to reality, even though he too, 
fails to provide the singularity free electron. To admit to singularities does not 
seem to me the right way. I think that in order to make progress we must once more 
find a general principle conforming more truly to nature.''} \cite{ein:94} \footnotemark[2]

\section{Klein's compactification}

In 1926 Oskar Klein returned to Kaluza's theory and wrote two classic papers \cite{kle:27,kle:26}. 
His more elaborate paper in the Zeitschrift f\"ur Physik is divided in a review of Kaluza's classical theory 
and the study of the connection between the quantum-mechanical wave-equation and five dimensional (null-)geodesics. In 
the first part Klein puts forward a more fruitful interpretation of the 
five-dimensional metric:
\be
g^{(5)}_{IJ}= \left( \begin{array}{rr}
g^{(4)}_{\mu \nu} + V A_{\mu} A_{\nu} &  V A_{\nu} \\
V A_{\mu} &  V
\end{array} \right) 
\label{eq:klee}
\ee
He continues by putting the dilaton $V$ to a constant, as this does not violate
four dimensional general covariance or the electro-magnetic gauge symmetry. 
Consequently he varies the action and finds the full field equations of general relativity, with the energy-momentum tensor of the electro-magnetic fields, and the 
source free Maxwell equations\footnote{The inconsistency in putting the dilaton to a constant \emph{after} varying, as noted by for instance Thiry \cite{thi:48}, thus 
does not surface. The field equation for the dilaton would then namely impose $F^2 =0$. For more on the role of the dilaton, see also \cite{str:98}.}:  
\be
\begin{array}{ll} G_{\mu \nu}=\kappa T_{\mu \nu}  \\
\nabla_{\mu}F^{\mu \nu}=0.  \end{array}      
\ee
So Kaluza's vacuum theory with the cylinder condition contains the full field 
equations.

Klein would continue by taking as a fact the quantized nature of the 
electric charge, which would lead him to the conclusion that the fifth direction 
needs to be compact. In his German article he explicitly studies a wave field, which he writes as
\be 
U = \exp ( i \hbar^{-1} \Phi(x^I)) 
\ee
 and for which
Klein in effect postulates the massless wave equation in five dimensions
\be 
\frac{1}{\sqrt{-g}} \p_{I} (\sqrt{-g} g^{IJ} \p_{J}) U =0.
\ee
With the identification 
\be 
p_{I }= \p_{I} \Phi
\ee
he deduces that in leading order ($\sim \hbar^{-2}$) the wave equation can be interpreted as a null geodesic in five dimensions: $p_{I}p^{I}=0$. This null geodesic is then reinterpreted as the geodesic for a charged particle in a four dimensional space, for which one has to put $p_{5}=q/c\sqrt{2\kappa}$ with $q$ the electric charge. 
Klein takes the value of the fundamental charge $e$ for $q$, arguably equivalent to writing the wave field $U$ as an eigenstate with eigenvalue $p_{5}=e/c\sqrt{2\kappa}$:
\be
U = \exp{(i e (\hbar c \sqrt{2\kappa})^{-1} x^5)}\,  \psi (x^{\mu}). \label{eq:schietop}
\ee
After performing the differentiation with respect to $x^5$, the wave-equation takes the 
familiar Klein-Gordon form: $D_{\mu} D^{\mu}\psi -m^2 \psi=0$ with $m=\frac{e}{\sqrt{2\kappa} c^2}$ and the gauge field $A_{\mu}$ incorporated in the covariant derivative in the usual way. One can reinterpret the particle's momentum in the fifth direction as a rest mass in four dimensions, since it moves along a five dimensional null geodesic and thus does not have a rest mass in five 
dimensions. The mass is of the order of the Planck mass. 
Klein realizes that the 
discreteness of the charge spectrum, via the de Broglie relation, leads to a 
discrete wavelength in the fifth direction: 
\be
p_{5}=\frac{\hbar}{\lambda_{5}}=\frac{ne}{(2\kappa)^{1/2}c}
\ee
This then tells the fifth direction needs to be compact, periodic with period 
$2\pi \lambda_{5}$. Putting in the values of the parameters tells the scale of the 
fifth direction is of about the Planck size: 
\be
\lambda_{5}=\frac{hc \sqrt{2 \kappa}}{e}=0.8 \times 10^{-30} \mbox{cm}. \label{eq:straal}
\ee 
\emph{``The small value of this length together with the periodicity in the fifth dimension may perhaps be taken as a support of the theory of Kaluza in the sense that they may explain the non-appearance of the fifth dimension in ordinary experiments as the result of averaging over the fifth dimension."} \cite{kle:27}\\
\\
With (\ref{eq:schietop}) one sees that translations $x^5 \rightarrow x^5 + \Lambda(x)$ correspond to gauge transformations of the wave-function. The gauge parameters take values on the circle, the gauge 
group of electrodynamics is now $U(1)$. The topology of space has changed as 
the fifth direction is no longer represented by an infinite straight line, that one can scale arbitrarily, but must be thought of as a circle, with a fixed scale. 

Einstein, shortly after Klein, published a paper that essentially reproduced Klein's 
results in the classical theory (which he acknowledges in an appendix), but did not make any mention of a possible compactification or the quantization of charge \cite{ein:27}. We see here implicitly that in unified field theory he did not start out by taking as hypothesis or axiom a fact from nature with which he would consequently deduce or construct - a method he had been successful with in the past. Rather, he expected the classical theory to produce the quantum results and he did not want to start out by using typical quantum relations.  

In a five dimensional theory of relativity in which one takes 
the fifth direction compact, one can expand the metric's components in a Fourier series:
\be 
g_{IK}=\sum_{n} g^{(n)}_{IK}(x^{\mu}) e^{inx^{5}/\lambda_{5}}. 
\label{eq:fourr}
\ee
This decomposition one generally finds in more recent literature (for instance \cite{duf:86}). In this form one can identify the charged sources with the higher order Fourier components of the metric. Klein 
certainly did not write the metric in the above form. However, his introduction of 
the wave field is to some extent equivalent: one can put 
$T^{(n)}_{IK}=(\p_{I}U^{(n)})^{*} \p_{K}U^{(n)}$ for waves representing charge 
$ne$, and identify these terms with the higher Fourier component contributions to 
the five dimensional Ricci tensor derived from the metric in (\ref{eq:fourr}). 
When pushing this analogy a bit further, one can show that the discreteness of charge is reproduced by a priori quantizing the field 
theory, i.e. turning continuous functions into counting operators. We would like 
to exemplify this: take a two dimensional 
space upon which the massless Klein-Gordon field $\phi$ lives, which we take to represent the (diagonal) components of the metric. The dependence on $x^5$ is of 
higher order:
\be
\phi(x^{0},x^{5})= e^{-im x^0} \sum_{n} A^{(n)} e^{inx^{5}/\lambda_{5}} \,\,\,\,\,\, \& 
\,\,\,\,\,\, g^{AB} \sim \left( \begin{array}{cc}
-(A^{(0)})^{-1} &      0 \\
0     		& (A^{(0)})^{-1}	
\end{array} \right) \label{eq:babs}
\ee
Charge is the component of momentum in the $x^5$ direction: 
\be 
P^{5}= \int dx^5 \,\,\, T^{05} = \int dx^5 g^{A0}g^{B5} (\p_{A}\phi)^{*} 
\p_{B}\phi
\ee
and using (\ref{eq:babs}) gives:
\be 
P^{5}= m (A^{(0)})^{-2} \sum_{n}n (A^{(n)})^{*} A^{(n)}. \label{eq:pvijf}
\ee
The frequency $m$ is interpreted as mass when going down one dimension 
(e.g. from five to four). But in the higher dimensional space the wave propagates 
on a null geodesic, and thus $c=m\lambda_{5}$. Using this relation one can rewrite the momentum in the fifth direction: 
\be
P^{5}=c/\lambda_{5}\, \times \, (A^{(0)})^{-2} \sum_{n}n (A^{(n)})^{*} A^{(n)}=const./\lambda_{5}.  \label{eq:pzes}
\ee
The coefficients $A^{(n)}$ are just complex numbers that can take a continuous range of 
values, so classically the compactification does not give a discrete spectrum for 
the charge, classically there is no minimum value for the 
constant. After quantization, the discrete charge spectrum does imply a compact fifth direction 
and vice versa. In other words, imposing the de Broglie relation with integer charges is consistent with turning the continuous constants $(A^{(n)})^{*}$ into creation operators.   

The tower of charged spin-2 particles one retrieves from the $5\times 5$ metric (\ref{eq:fourr}) upon 
quantization have very high masses $m_{n} = \frac{ne}{\sqrt{2\kappa}c^2}$ 
(their gravitational interaction would be of the same order of magnitude as the electro-magnetic interaction). As 
effective theory for low energy physics, one can retain just the $n=0$ mode in the 
Fourier expansion. This gives for the action: 
\be
 R^{(0)(4)}+ \frac{1}{4} V^{(0)} F^{(0)}_{\mu \nu}F^{(0)\mu \nu} 
-\frac{2}{V^{(0)1/2}} \Box V^{(0)1/2}  
\ee 
which is just the action of the theory with the cylinder condition in place (referred to as \emph{Kaluza's theory}, the theory with higher Fourier components -in other words, with periodic $x^5$ dependence- will be called \emph{Kaluza-Klein theory}). 
Note that the Kaluza theory has forgotten about the original periodicity in $x^5$ \cite{duf:86}.

\section{Einstein on classical Kaluza-Klein}

We now make a leap in time and turn to the study Einstein made of classical 
Kaluza-Klein theory in the years 1938-1943. 
The first paper of this period 
starts with a reevaluation of Kaluza's theory, where the dilaton has been set to a constant (thus $L=\sqrt{-g}(R+cF^2)$). Einstein, together with Peter Bergmann, remarks the following:\\
\\
\emph{"Many fruitless efforts to find a field representation of matter free from singularities based on this theory have convinced us, however, that such a 
solution does not exist.} footnote: \emph{We tried to find a rigorous solution of the gravitational equations, free from singularities, by taking into account the 
electro-magnetic field. We thought that a solution of a rotation symmetric 
character could, perhaps, represent an elementary particle. Our investigation was 
based on the theory of 'bridges'...We convinced ourselves, however, that no 
solution of this character exists.'' } \cite{ein:38}\\
\\
This makes it very plausible that, as Einstein had not been able to find a 
non-singular object he sought to expand 
the contents of the theory he studied\footnote{Possibly, Einstein and Bergmann did not succeed in formulating the modification of the field equations as proposed by Einstein and Rosen (i.e. $g^2 (R_{\mu\nu}+T_{\mu\nu}=0$) in the context of Kaluza's theory. This modified form of the field equations was essential in the interpretation of the Einstein-Rosen bridge as a non-singular object (see, also with regard to Einstein's handling of the Schwarzschild singularity, \cite{ear:99}).}.
Einstein and Bergmann now proposed a theory 
in which the cylinder condition had been relaxed: (periodic) dependence on $x^5$ would again be allowed. They do note, however, that spacetime appears to us as four 
dimensional and they explain this as follows (with a two dimensional analogy, see also figure 1): \\
\begin{figure}
\begin{center}
\includegraphics[height=2.5in, width=4in]{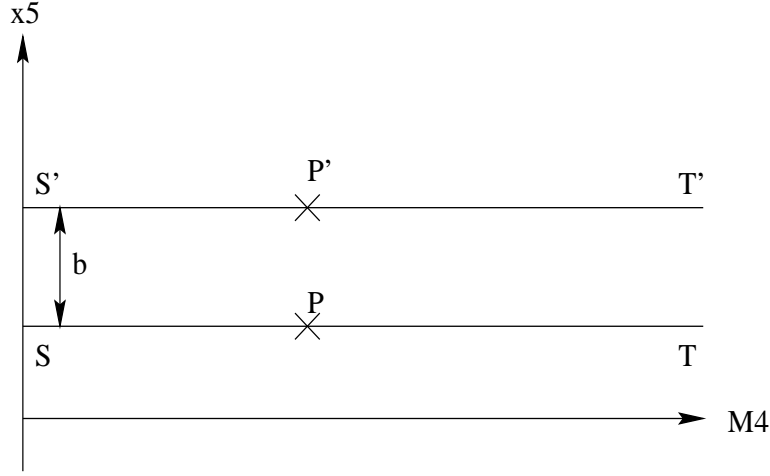}
\caption{\small Einstein and Bergmann gave the above 2 dimensional image of the compactified space they studied:\emph{``We imagine the strip curved into a tube so that ST coincides with S'T'.''} \cite{ein:38}} 
\end{center}
\end{figure}
\\
\emph{``If the width of the strip, that is the circumference of the cylinder 
(denoted by b), is small, and if a continuous and slowly varying function 
$\phi(x^{0},x^1 )$ is given, that is if $b \p\phi /\p x^a$ is small compared with 
$\phi$, then the values of $\phi$ belonging to the points on the segment PP'} 
[i.e. the line connecting the periodically identified points P and P'] 
\emph{differ from each other very slightly and $\phi$ can be regarded, 
approximately, as a function of $x^1$ only.''} \cite{ein:38}\\
\\    
Not only do Einstein and Bergmann take the scale of the extra and compact direction to be 
small, they also make the tacit assumption that the $n\neq 0$ Fourier coefficients ($A^{(n)}$) are 
small relative to the $n=0$ coefficient\footnote{Einstein 
and Bergmann did not explicitly write the Fourier expansion of the metric.}. 
However, they cannot, in their classical theory, explicitly calculate the
value of the scale parameter $b=\lambda_{5}$ from the electron charge. They do not a priori quantize and therefore cannot use the value of $\hbar$, like Klein did when using the de Broglie relation.  

By relaxing the cylinder condition, it is again possible to have classical fields 
in the theory that carry a momentum in the fifth direction. These then represent 
charged matter sources, which were absent in Kaluza's theory with constant dilaton. 
Einstein and 
Bergmann find the following field equations (with $g_{55}=const$):
\be \nonumber
\alpha_{1}(R_{\mu \nu}-\frac{1}{2}g_{\mu\nu}R)+ \alpha_{2}(2F_{\mu \alpha}
F^{\alpha}_{\nu}-\frac{1}{2}g_{\mu\nu}F^{\alpha\beta}F_{\alpha\beta}) +\\ 
\nonumber
+\alpha_{3}(-2g_{\mu\nu,55}+2g^{\alpha\beta}g_{\mu\alpha,5}g_{\nu\beta,5}
-g^{\alpha \beta}g_{\alpha\beta,5}g_{\mu\nu,5}-\frac{1}{2}g^{\alpha\beta}_{,5}
g_{\alpha\beta,5}g_{\mu\nu}) +\\ 
+\alpha_{4}g_{\mu\nu}(\frac{1}{2}(g^{\alpha\beta}g_{\alpha\beta,5})^2
+2g^{\alpha\beta}g_{\alpha\beta,55}+2g^{\alpha\beta}_{,5}g_{\alpha\beta,5})=0  \label{eq:genein}
\\ \nonumber
\\ 
\int dx^5 \, \sqrt{-g}\,\,\, 
\Big\{ \alpha_{1}(g^{\alpha \beta}\Gamma^{\mu}_{\alpha \beta,5}-g^{\alpha \mu}
\Gamma^{\beta}_{\alpha\beta,5})-4\alpha_{2}F^{\mu\alpha}_{;\alpha} \Big\} =0.
\label{eq:jamax} 
\ee
One sees in the last expression (\ref{eq:jamax}), i.e. Maxwell's 
equations, very clearly that the $x^5$-dependent metric fields play the role of charged matter sources. 
The $\alpha_{i}$ are coupling constants, that in this paper are still arbitrary 
because in this version of the theory five dimensional general covariance is not 
imposed. In a second paper on this theory they do impose general covariance which 
fixes the constants \cite{ein:41}. (Einstein, Bergmann and, by then joined by, 
Valentin 
Bargmann are at first uncomfortable about the arbitrariness in the value of these 
constants \cite{ein:95}. When this arbitrariness 
is lifted by general covariance one faces a new problem because then the 
electro-magnetic and gravitational interaction are of the same order of magnitude 
\cite{ein:41}.)

Now, why did indeed Einstein set out to study this generalization of Kaluza's five 
dimensional theory? What made him study Kaluza-Klein as a classical field theory? 
To formulate a unified theory, just because a 
unification of forces would have such appeal? Now that he had formulated field equations which contain geometrically construed source terms - what were his ensuing goals?

In our search for an 
answer to these questions we have looked at the correspondence Einstein had with 
Bergmann, Bargmann and also Wolfgang Pauli. Firstly, one learns that Einstein was quite happy with the results;\\
\\
\emph{``It is a pity that our paper starts with Kaluza ... But this is just 
a minor detail; I have a lot of confidence in the theory as such.''} \cite{ein:98} \footnotemark[2]\\
\\
Bergmann writes:\\
\emph{``Regardless of the difficulties that have surfaced I too very much confide 
in the theory because of its unity and the existence of `wave type' 
fields.''} \cite{ber:38} \footnotemark[2]  \\
\\
It is enticing to wonder how Einstein would hope to undercut quantum theory with classical Kaluza-Klein theory. 
One expects this to be the primary objective for Einstein to study this theory. However, in none of the papers nor in his 
correspondence is their any explicit mention of how this should come about. At 
certain points in the discussion with Bergmann, there is nevertheless mention of 
a ``de Broglie frequency''. But it seems the few relations that are introduced 
are done so merely as consistency check, for example, to see if the equations 
allow solutions with negative masses. We will, however, later return to 
this point to see in what ways Einstein may have wanted to derive typical quantum results. 
The wave field Bergmann refers to above probably corresponds to a 
field that has a wave propagating in the fifth direction, giving an electrically 
charged source term in the classical field equations.\footnote{\emph{``We had put... $\sim \gamma_{\mu\nu} = 
\psi_{\mu\nu}e^{2i\pi x^5 /\lambda} + \bar{\psi}_{\mu\nu} e^{-2i\pi x^5 
/\lambda}=$ real quantities, bar means complex conjugate...for our stationary problem...$\psi_{\mu\nu}=\chi_{\mu\nu}e^{i \omega t}$.''} 
Taken from \cite{ber:37}, our translation.} \\  
\\
In September 1938, Pauli writes Einstein:\\
\emph{``Bargmann}\footnote{Valentin Bargmann joined Bergmann and Einstein in their work on Kaluza-Klein theory around this time.} \emph{was recently here and reported on your work on the closed 
5-dimensional continuum. But that is an old idea of O. Klein, including the 
circumstance that the $\psi$-function gets a factor of $e^{ix_{5}}$ and he has 
always emphasized that changing the origin of $x_{5}$ corresponds to so called 
gauge group transformations. Besides our old, more fundamental differences of 
opinion, your Ansatz appears to me too special, because it is hardly necessary for 
the $\psi$-function to be a symmetric tensor.''} \cite{pau:38} \footnotemark[2]\\
\\
Pauli continues by saying that in his opinion, Bargmann is more talented in 
mathematics than in physics. Then:\\
\emph{``In this context I would like to end with a little malicious remark: considering your current involvement with theoretical physics, this should hardly make a bad impression on you.''} \cite{pau:38} \footnotemark[2]\\
\\
Einstein replies: \\
\emph{``Your remark is well founded, and could easily be answered likewise.}\footnote{Possibly Einstein is referring here to the work Pauli had done on the five dimensional theory in 1933 \cite{pau:33,pau:34}.} \emph{But the new work is only superficially similar to Klein's. It is simply a logical improvement of Kaluza's idea, that deserves to be taken serious and examined accurately.''} \cite{ein:99} \footnotemark[2]

\subsection{\emph{``Ich hoffe dass dies nun glatt gehen wird, ohne dass uns die 
ewige G\"ottin erneut die Zunge herausstrecken wird.''} \cite{ein:97}}

From his correspondence one learns how Einstein wanted to test the 
generalization of Kaluza's theory: he wanted to find a non-singular charged 
particle. Just after having written the paper with Bergmann, published in the 
Annals of Mathematics in July of 1938, Einstein started his attempts at 
realizing this non-singular object. The discussion lasted for at least a 
year. One equation under consideration was the following: \\
\\
Bergmann to Einstein, August '38\\
\emph{''...in our case the lowest power ... is: 
\be 
 +\frac{2\pi}{\lambda}\omega[ \alpha (3\alpha +2\beta)]=\alpha_{2}(\phi'' + 
\frac{2}{\rho}\phi') \label{eq:poiss}
\ee
 i.e. the sign of the charge (term on the right hand side) and of $\omega/\lambda$ 
also depends on the sign of the square bracket which is different for both 
particles.''} \cite{ber:37} \footnotemark[2]\\
\\
$\lambda$ is the scale of the fifth direction, $\omega$ the frequency of a wave 
traveling in the fifth direction. $\alpha_{2}$ is a coupling constant, $\rho$ is a 
radial coordinate ($\rho^2=r^2 +a^2$ with $a$ a constant and $r$ the usual radial coordinate). $\phi$ is the 
electrostatic potential which is differentiated with respect to $\rho$. This is the 0-component of the Maxwell equations 
(\ref{eq:jamax}), i.e. the Poisson equation, written in the leading terms at large distances. The solution of the full equation
would correspond to an object represented by a wave propagating in the fifth direction and curving the 4-dimensional spacetime.
$\alpha$ and $\beta$ stem from the four dimensional curved metric.
One then has 
a body sitting somewhere in space, with a momentum in the fifth direction that 
yields the electric charge.  

What type of exact solution they hope to retrieve can be read in for instance the 
following note from Einstein to Bergmann: \\
\\
\emph{``The treatment of our centrally symmetric problem is still mysterious to me. 
We in any case need a factor that is essentially singular at infinity... Maybe we 
will have to introduce $\rho$ again. But I still do not have a comprehensive view 
of the situation, particularly as $\phi$ has to go as $\epsilon/r$. One would need 
to know a lot about differential equations to find this solution. Perhaps Bargmann 
will find a way. It is certainly not easy. Anyhow, one first needs to find a 
representation at large $r$ and then strive for a convergent representation at the 
origin.''} \cite{ein:98} \footnotemark[2]\\
\\
One sees that Einstein wants a particular solution that drops off as $\epsilon / r$ at large distances; as one is far away from the object one 
should see an electric monopole charge sitting somewhere in space. Indeed, we 
would like to emphasize Einstein wants a non-singular, convergent series for the 
potential at the location of the source. (At some point, they expected that 
substituting $\rho$ instead of $r$ would give them a convergent series.) 

In case of the charged wave solution one could allow the relation $E= h \nu$, and with the de Broglie relation the five dimensional photon carries a charge given by $p_{5}=\frac{h}{\lambda}$. It is possible this consideration would be related to the following remark by Einstein:\\
\\
\emph{``We have demanded that the frequency $\nu$ is equal to a de Broglie frequency. So far, this had only been supported by physical plausibility, not founded on the mathematical problem. However, it has now become clear to me this choice will have to be made.''} \cite{ein:96} \footnotemark[2]\\
\\
The quantized nature of electric charge can be in accordance with the compactification of the fifth direction. But then one has the exact line of reasoning Klein followed. However, that is a priori only conclusive once one makes the transition from classical fields to quantum fields, i.e. from numbers to operators: without quantizing, $P_{5}$ of the wave-field would have a continuous range of values: $const/\lambda = m \sum_{n}n(A^{(n)})^{*}A^{(n)}(A^{(0)})^{-2}$ (eqn. (\ref{eq:pzes})). So without quantization the compactification looses its explanatory power regarding the discreteness of the charge spectrum. As Einstein nowhere in his correspondence, nor in his articles, assumes a discrete charge spectrum it becomes less probable that one should interpret the above remark as giving in to Klein's reasoning. 

It is more likely that he hoped that the generalized Poisson equation (\ref{eq:poiss}), with the boundary conditions he imposed and combined with the generalized Einstein equations (\ref{eq:genein}), would produce a discrete spectrum of charged solutions. That is to say, that the charges $P_{5}$
take on a discrete range of values because the field equations and boundary conditions would restrict the integral to a discrete set of values. Note that because of the compactification, there would then be a fundamental constant in the theory with the dimension of angular momentum:
\be 
P_{5} \, \lambda \, = \, h.
\ee
One could set this to the value of Planck's constant and arrive at a value for $\lambda$. Then there would not be some ad hoc quantization implemented. Instead, the discretely charged solutions of the classical field equations, together with the compactification radius would yield Planck's constant. So there would not be an arbitrary quantization invoked, but the geometric scheme -compactification- and the regularity conditions would now produce the existence of a minimal quantum of action $h$. Yet, we emphasize that there is no explicit indication in either his correspondence or his articles that the quantization of charge is an actual result he is looking for, and that the above argument would be his method for rederiving $h$.

For at least a year Einstein, Bergmann and Bargmann try to find a non-singular 
charged object in the classical Kaluza-Klein theory (with constant dilaton). Finally, one can read in their 
joint paper (that came out in 1941 but was probably written sooner, as the slightly 
different approach -basing the theory on five dimensional general covariance- laid 
out in this work is already mentioned in a letter form Einstein in 1939 \cite{ein:95}) that this 
search had become frustrated: \\
\\
\emph{``It seems impossible to describe particles by non-singular solutions of the 
field equations. As no arbitrary constants occur in the equations, the theory would lead to electro-magnetic and gravitational fields of the same order of magnitude. Therefore one would be unable to explain the empirical fact that the electrostatic force between two particles is so much stronger than the gravitational force. This means that a consistent theory of matter could not be based on these equations.''} \cite{ein:41}

\section{Solitons in Kaluza's theory}

We have seen the emphasis Einstein put on non-singular particle descriptions (see also \cite{ear:99}), in particular in classical Kaluza-Klein theory. In 1941 he would publish a short paper on the non-existence of non-singular massive objects in four dimensional relativity theory \cite{ein:42}. The argument would later be extended to Kaluza's theory in a joint publication of Einstein and Pauli \cite{ein:43}. They argue that solitons cannot exist in this theory, a result that in particular Einstein may have been disappointed with. Nevertheless, in the 1980's solitons were described in the Kaluza-theory. We will briefly discuss these, and compare them with the Einstein-Pauli theorem.

\subsection{Non-singular particles and Noether currents in relativity}

We briefly discuss Einstein's argument for the non-existence of non-singular particle solutions of the vacuum field equations of four dimensional relativity. 
  This theory contains the conserved current $\aleph^{\mu}$:
\be 
\aleph^{\mu}_{,\mu}=\Big( \sqrt{-g} [g^{\alpha \mu} \delta \Gamma^{\beta}_{\alpha \beta}   - g^{\alpha \beta} \delta \Gamma^{\mu}_{\alpha \beta}]\Big)_{,\mu}=0. \label{eq:tucum}
\ee
Einstein in his paper \cite{ein:42} derives the conservation law (\ref{eq:tucum}) by considering the variation $\delta R_{\mu\nu} = -(\delta \Gamma^{\alpha}_{\mu\nu})_{;\alpha} + (\delta \Gamma^{\alpha}_{\mu\alpha})_{;\nu}$ which he rewrites as:
\be 
\sqrt{-g}g^{\mu\nu}\delta R_{\mu\nu}=\aleph^{\alpha}_{,\alpha} \label{eq:vara}
\ee
with $g_{\mu\nu}+\delta g_{\mu\nu}$ the variation of a line-element of an arbitrary, everywhere regular field. He then assumes that both the varied and original fields satisfy the vacuum field equations throughout the spacetime ($R_{\mu\nu}=0$), from which it follows that the variation (\ref{eq:vara}) should vanish. Finally, the integral over spacetime of (\ref{eq:tucum}) should vanish for non-singular spaces. 

Einstein's argument is proposed to hold under the following quite general conditions: \\
\\
\bfseries{Assumption A}: \mdseries{\emph{``Let us consider now a solution without singularities of the gravitational equations plunged in an Euclidean (or Minkowski) space, which we assume either to be independent of $x^4$ or periodic or quasi periodic with respect to $x^4$. Such a solution would be a theoretical representation of a body, or, respectively a system of bodies, which in the average is at rest with respect to the coordinate system. At great distances from the origin of coordinates, the field of such a system may always be replaced by that of a resting point mass...''} \cite{ein:42}}\\
\\
Asymptotically $g_{\mu\nu}=\eta_{\mu\nu}+\gamma_{\mu\nu}$ for which the linearized field equations give: 
\be 
\gamma_{ik}=-\frac{2m}{r}\delta_{ik}, \,\,\,\,\,\,\,\,\,\,\,\,\, \gamma_{44}=-\frac{2m}{r}. \label{eq:oppll}
\ee
The leading term in the variations will be
\be 
\aleph^{i}=-2(\delta m) \eta^{ik}\Big(\frac{1}{r}\Big)_{,k}, \,\,\,\,\,\,\,\, \aleph^{4}=0.
\ee
One finds with Einstein's assumption for the time-dependence and the assumption of regularity ($T$ is time):
\be 
\int d^4 x\,\, \aleph^{\mu}_{,\mu}=\int dt \int d^2 \Omega_{i} \,\,\aleph^{i}=8\pi T\delta m  =0
\ee
\emph{``Result: two infinitely close solutions without singularities necessarily have the same mass''} \cite{ein:42}. On the other hand, if (\ref{eq:oppll}) solves the equations of motion, then on the basis of general covariance there should also be a solution for $r$ replaced by $(1+c)^{-1} r$, i.e. a non-singular solution with mass $(1+c)m$, where one can take $c$ infinitely close to 0. So there should be non-singular solutions with infinitely close masses. Apparently, we have here a contradiction in terms: \emph{``This contradiction can only arise from the inexactitude of the hypothesis that there exists a solution free of singularities belonging to a total mass different from zero''} \cite{ein:42}\footnote{Andr\'e Lichnerowicz generalized Einstein's result in 1946 \cite{lic:46}, see also \cite{lic:55}. Einstein's result tells that there can be no masses (identified as the constant that multiplies the $1/r$ term in the metric) in non-singular spaces, Lichnerowicz showed that higher order multipole-terms cannot be present either. 
He could show that, relying on Einstein's argument, a non-singular space with \emph{Minkowskian} signature necessarily has a trivial time-component and consequently the curvature should be realized in the three space-like dimensions. But in three dimensions, if $R_{ij}=0$ then $R^{k}_{ijl}=0$ and the three manifold is flat.
Actually, Lichnerowicz had not seen Einstein's publication (it was not available in France \cite{lic:45}), but he deduced its argument from its five dimensional analogon in the paper Einstein and Pauli published in 1943. Lichnerowicz had formulated a similar singularity theorem in his thesis \cite{lic:38}, repeated in a publication \cite{lic:39} and had sent this to Einstein already in 1939 \cite{lic:45}, but it had not arrived \cite{tip:80}. In 1945 Lichnerowicz corresponded with Pauli on his generalization and informed both Einstein and Pauli of his work.}.

\subsection{Einstein and Pauli on non-singular solutions in Kaluza's theory}

Einstein's argument would be repeated, by himself and Pauli in an article on Kaluza's theory. So, the theory in which they reformulate the argument is the vacuum Kaluza-Klein theory where the higher modes have been dropped, equivalent to a five dimensional theory of relativity with a Killing vector $\p/\p x^{5}$ where a priori the extra dimension need not be compact. They start of as follows:\\
\\
\emph{``...whether in a five dimensional metric continuum (of signature 1) the equations $R_{IK}=0$ admit of non-singular stationary solutions with a field $g_{IK}$ asymptotically given by}  
\be
\begin{array}{rrrrr}
B & 0 & 0 & 0 & D \\
0 & A & 0 & 0 & 0 \\
0 & 0 & A & 0 & 0 \\
0 & 0 & 0 & A & 0 \\
D & 0 & 0 & 0 & C
\end{array} \label{eq:asympt}
\ee
\emph{where at least one of the quantities A,B,C,D has the form $\pm 1 + \frac{const}{r}$ with a non-vanishing constant. This is the asymptotic form of a field representing a particle whose electric and ponder-able masses do not both vanish.''}\footnote{Note that introducing a Coulomb field in $g_{05}$ would lead to a term proportional to $1/r^2$ in $A$ and $B$ as in the Reissner-Nordst\"om metric. However, this term can be disregarded at large distances, as it is of higher order.} \cite{ein:43}\\
\\
Einstein and Pauli do not demand spherical symmetry. Note that for the first time since Klein, again the corner component $g_{55}$ is retained. The possible non-constancy of this component and its interpretation as a 'variable' gravitational constant would be emphasized by Jordan, and would have been published by him in 1945, but then the Physikalische Zeitschrift had ceased publication. We learn from an article by Bergmann that Pauli at that time knew about this work and he further informs us that he himself and Einstein had considered the possibility as well, \emph{``several years earlier''} \cite{ber:48}. Then, it is perhaps not that unexpected that this component would again be considered a variable by Einstein and Pauli in 1943.\footnote{Also Thiry in 1948 would reintroduce the $g_{55}$ component as a variable \cite{thi:48}. He mentions that Lichnerowicz's theorems should apply to the resulting theory, see also \cite{lic:55}. For Jordan's work, see \cite{jor:47}.}  

Einstein and Pauli reestablish relation (\ref{eq:vara}), but now in five dimensions:
\be 
\aleph^{S}_{,S}=0
\ee
with $\aleph^{S}$ as in (\ref{eq:tucum}) for the metric of the five dimensional space. One imposes time-independence and the cylinder condition:
\be 
\frac{\p g_{IK}}{\p x^\ep} =0 \,\,\,\,\,\,\,\,\,\,\,\, \ep =0,5 \label{eq:epc}
\ee
With Gauss' theorem and (\ref{eq:epc}) one finds for non-singular spaces:
\be  
\oint \aleph^{i}n_{i}d^2 \Omega =0  \label{eq:oint}
\ee
For singular spaces one has:
\be  
\oint_{F_{1}} \aleph^{i}n_{i}d^2 \Omega=\oint_{F_{2}} \aleph^{i}n_{i}d^2 \Omega
\ee
where $F_{1}$ is the inner surface surrounding the singularity and $F_{2}$ is an outer boundary. The variations are performed via coordinate transformations ($x^I \rightarrow x^i + \xi^I$), that leave (\ref{eq:epc}) invariant. There are two types of these variations. The first is independent of the coordinates $x^{\ep}$ with $\ep=0,5$. These, and its derivatives can be set to zero on the inner surface $F_{1}$, and on the basis of (\ref{eq:oint}) one would not be able to decide whether the space is singular or not. The other type is given by;
\be 
\xi^i=0, \,\,\,\,\,\,\, \xi^{\ep}=c_{\rho}^{\ep} \xi^{\rho}\,\,\,\,\,\,\,\,\,\,\,\,\,\,
(i=1,2,3 \,\,\,\,\,\, \ep, \rho=0,5)
\ee
with constant coefficients $c^{\ep}_{\rho}$. With these transformations (\ref{eq:oint}) would single out non-singular solutions. By calculating the variations $\delta \Gamma^{I}_{AB}$ the condition for non-singular spaces reduces to:
\be 
c^{\ep}_{\rho} \oint \sqrt{|g^{5}|}g^{\rho A} \Gamma^{i}_{\ep A} n_{i}d^2 \Omega=0.
\ee
One can drop the $c^{\ep}_{\rho}$ and the summation implied, as the constants can be chosen arbitrary anyway. Again solving the linearized field equations for the asymptotic fields (\ref{eq:asympt}) yields;
\be
g_{IK} =
\left( \begin{array}{rrrrr}
-1+\frac{m_{00}}{r} & 0 & 0 & 0 & \frac{m_{05}}{r} \\
0 & 1+\frac{m}{r} & 0 & 0 & 0 \\
0 & 0 & 1+\frac{m}{r} & 0 & 0 \\
0 & 0 & 0 & 1+\frac{m}{r} & 0 \\
\frac{m_{05}}{r} & 0 & 0 & 0 & 1+\frac{m_{55}}{r}
\end{array} \right) \label{eq:sy}\\
\nonumber \\
\mbox{as}\,\,\,\, r \rightarrow \infty \,\,\,\,\,\,\,\mbox{and} \,\,\,\, m+m_{55}=m_{00}. \label{eq:demd}
\ee
Putting this in the condition (\ref{eq:oint}) leads to 
\be
\oint \sqrt{|g|}g^{\rho A} \Gamma^{i}_{\ep A }n_{i}df=\pm 2 \pi m_{\rho \ep}.
\label{eq:result}
\ee
For non-singular spaces one finds:
\be  
m_{\rho \ep}=0
\ee
and Einstein and Pauli conclude the only non-singular solution of the vacuum field equations is five dimensional Minkowski space.

\subsection{Two examples of solitons in Kaluza's theory}

In the 1980's solitons have been described in the vacuum Kaluza-theory \cite{gro:83,sor:83}. Generally, these are instantons from Euclidean gravity, incorporated in Kaluza's theory in the metric's components in the space-like directions, and for the remainder one takes $g_{0A}=-\delta_{0A}$. We describe two of these objects in more detail here and we discuss why the Einstein-Pauli proof does not apply to them. We will disregard the $x^5$ dependent higher Fourier modes of the Kaluza-Klein theory, so we are in the $n=0$ Kaluza theory as Einstein and Pauli. If one has a compact dimension, this is a principal bundle of $U(1)$ over the four dimensional spacetime. In the $n=0$ theory, as argued before, there may not be an a priori reason for the theory to have a compact extra dimension. We will first study the magnetic monopole in this theory, that by virtue of a regularity condition imposes the fifth dimension to be compact
and is an example of a non-trivial bundle. Basically it is the Dirac monopole \cite{dir:31} in five dimensions. In fact, it is a generalization of the Euclidean Taub-NUT (see \cite{mis:63, haw:77}) space to five dimensions, and thus  the Kaluza-Klein theory shows the equivalence of these two objects. For both objects the topological $S^3$-structure and its consequential description with the Hopf-map had already been uncovered (for the Dirac monopole this was done by Wu and Yang \cite{wuy:75} and in case of the Taub-NUT space by Misner \cite{mis:63}. For an overview, see \cite{bai:83}). 

A natural restriction one can put to the $n=0$ theory when one is looking for solitons that are intended to represent particle solutions, is to demand the solutions to be static. Einstein and Pauli also put forward this restriction in their theorem on particle solutions and one did the same in the eighties, when Kaluza-Klein theory was again intensively studied. This reduces the field equations to the equations of Euclidean gravity on surfaces of constant time, and the fifth dimension now plays the same role as the Euclidean time in the four dimensional theory \cite{gro:83}. With the latter theory a number of gravitational instantons had been described (by Hawking et.al. \cite{haw:77, gib:79}). Consequently, these were incorporated as solitons in the five dimensional Kaluza theory by adding in a topologically trivial way the time-coordinate to the instanton metric. This was carried out for the monopole by Gross and Perry \cite{gro:83} and Sorkin \cite{sor:83}, and for a number of other solitonic objects also by Gross and Perry \cite{gro:83}. 

\subsubsection{Gravitational nut and Kaluza-Klein monopole}

The gravitational instanton is defined by Stephen Hawking as follows (analogous to its definition in Yang-Mills theory): \emph{"a gravitational instanton..}[is]..\emph{a solution of the classical field equations which is non-singular on some section of complexified spacetime and in which the curvature tensor dies away at large distances"} \cite{haw:77}. Our first example is the Euclidean Taub-NUT gravitational instanton, a solution of the vacuum Euclidean Einstein equations. Its metric is;
\be
ds^2 = V (d\tau +4m (1-\cos{\theta})d\phi)^2+\frac{1}{V}(dr^2+r^2 (d\theta^2 + \sin{\theta}^2 d\phi^2))\\
\frac{1}{V}=1+\frac{4m}{r}
\label{eq:taubnut}
\ee
Reminiscent of the Dirac-string singularity, this metric is singular when $\theta=\pi$;
\be
g_{\tau \phi}= VA_{\phi}; \,\,\,\,\,\,\,\, A_{\phi}= 4m \Big( \frac{1-\cos{\theta}}{r\sin{\theta}}\Big).
\label{eq:vecpot}
\ee
One can remove this singularity by introducing two coordinate patches that are related to one another via the coordinate transformation \cite{mis:63}:
\be
\tau_{N}=\tau_{S}-8m\phi  \label{eq:ptau}
\ee
$A^{N}_{\phi}=g_{\phi \tau_{N}}/V$ transforms in the following way: 
\be
A^{S}_{\phi}=A^{N}_{\phi} - \frac{1}{r \sin{\theta}} \frac{\p (8m \phi)}{\p \phi}, \,\,\,\,\,\,\,\,\,  \mbox{so} \,\,\,\,\,\,\,\,\,
A^{S}_{\phi}= 4m \Big( \frac{-1-\cos{\theta}}{r\sin{\theta}}\Big),
\ee
 which is singular at $\ta = 0$. The singularities in the $A_{\phi}$'s are then obviously coordinate singularities and one can restrict oneself to the $(\tau_{N}, A^{N}_{\phi})$-description on the northern hemisphere (i.e. $\forall \, \ta$ except $\ta =\pi$) and to the $(\tau_{S}, A^{S}_{\phi})$-description on the southern hemisphere (i.e. $\forall \, \ta$ except $\ta =0$). One has with the two patches combined a complete non-singular description of the (Euclidean) Taub-NUT spacetime. Where the two patches overlap they are related by (\ref{eq:ptau}). Because of this condition $\tau$ must now be considered a periodic coordinate with period $16\pi m$, for $\phi$ is periodic with $2\pi$ \cite{mis:63}. In the Kaluza theory, this same regularity condition imposes the fifth dimension to be compact.

The Taub-NUT space has topology $S^3 \times R$, where $R$ represents the non-compact radial dimension. Its curvature goes to zero at large distances \cite{mis:63}, but as its topology at infinity is not the trivial bundle (i.e. not $S^1 \times S^2$) we cannot refer to the object as asymptotically Minkowskian. To show that the Taub-NUT instanton is regular at $r=0$ one can rewrite the metric with $r=\frac{1}{16m}\rho^2$: 
\be 
ds^2= \frac{\rho^2}{\rho^{2} + 64m^2}(d\tau +4m(1- \cos{\theta})d\phi )^2+\frac{\rho^2 + 64m^2}{64m^2}(d\rho^{2} +\frac{\rho^2}{4}(d\theta^{2} + \sin{\theta}^{2} d\phi^{2}))
\ee 
and in this form it is clear there is no singularity as $r \rightarrow 0$: the degeneracy at $r=0$ is analogous to the degeneracy in polar coordinates and is superfluous if one periodically identifies $\tau$ with period $16\pi m$ (the fixed point of $\p/\p \tau$ defines what is called a 'nut' \cite{gib:79}). It is clear the object satisfies Hawking's definition of an instanton. One can include the object as a soliton in Kaluza's theory, in the manner argued  before, namely by letting $x^5$ play the role of the Euclidean time $\tau$ and introducing a trivial time dependence. This then yields:
\be 
ds^2= -dt^2 + V(dx^{5} +4m (1-\cos{\theta})d\phi)^2+\frac{1}{V}(dr^2+r^2 (d\theta^2 + \sin{\theta}^2 d\phi^2))\\
\frac{1}{V}=1+\frac{4m}{r} \label{eq:gpmonop} 
\ee
The result is a magnetic monopole: $A_{\phi}$ has regained its interpretation as a component of the electro-magnetic vector-potential as in (\ref{eq:klee}), and the magnetic field is given by:
\be 
\vec{B}=\vec{\nabla} \times \vec{A}= \frac{4m \vec{r}}{r^3}
\ee
One hereby sees that the magnetic monopole charge $m$ is quantized ($m = \frac{1}{8}kR$, with integer valued $k$), as the transition function in the fiber takes the value $e^{i\frac{8m}{R}\phi}$ which has to be single valued on the equator\footnote{We had absorbed Kaluza's $\alpha = \sqrt{2 \kappa }$ in $\vec{A}$, rewriting it yields a reparametrisation: $B \rightarrow \alpha B$. Allowing the de Broglie relation Klein uses, together with (\ref{eq:straal}), would give the Dirac quantization condition: ($m=k \frac{hc}{e}$) and thus explain the quantization of electric charge. But classically (i.e. without de Broglie relation) the magnetic monopole tells nothing about the quantization of electric charge (one can have continuous electric fields in the background of the monopole spacetime).}. Most importantly for our story, $x^5$ transforms as:
\be
x^{5} \rightarrow x^{5} + 8 m \phi.
\ee
So the condition of regularity on the monopole spacetime induces the fifth dimension to be compact, with period $16\pi m$: it induces the fiber of the $n=0$ $x^5$-independent Kaluza theory to have the structure of $U(1)$, whereby the object reveals itself as a non-trivial bundle.

\subsubsection{Gravitational bolt as soliton}

The topology (i.e. the Hopf-fibering) of the Taub-NUT does not have asymptotically a trivial Minkowskian structure. This is at least one point where the monopole misses the connection with Einstein's theorems. In both papers (i.e. the paper with Pauli, and also the earlier paper with Grommer \cite{ein:43,ein:22}) on particle solutions in Kaluza's theory, Einstein focuses on objects that are not just Ricci flat at infinity, but asymptotically have the structure of Minkowski space. In what follows, we will give one more example of a $n=0$ Kaluza soliton, one that has the right asymptotic behavior.  In fact, it would fulfill all demands Einstein and Pauli put on particle solutions, all but one.  

Again, this Kaluza soliton can be thought of as a trivial extension of a particular gravitational instanton. Now we are dealing with the 'bolt' type instanton, which is just the Schwarzschild solution continued to the complex plane \cite{haw:77,haw:78, gib:79}:
\be 
ds^2 = (1-4m/r) d\tau^2 + (1-4m/r)^{-1} dr^2 + r^2 d\Omega^2. 
\label{eq:bolt}
\ee
This is a positive definite metric for $r > 4m$. The intrinsic singularity from Schwarzschild space, located at $r=0$, is excised from the Euclidean space. There is an apparent singularity at $r=4m$ but this is like the apparent singularity at the origin of polar coordinates, as becomes clear upon introducing a new radial coordinate $x=8m (1-4m r^{-1})^{1/2}$. The metric then becomes:
\be 
ds^2 = \Big(\frac{x}{8m}\Big)^2 d\tau^2 + \Big( \frac{r^2}{16m^2}\Big)^2 dx^2 + r^2 d\Omega^2
\ee 
If $\tau$ is regarded as an angular coordinate with period $16\pi m$ the above metric will be regular at $x=0, r=4m$. The corresponding regular coordinate system 
 of this space would be the Kruskal system \footnote{Writing the metric in the Kruskal form immediately shows the periodicity in $\tau$. Also, it again explains why the $r=0$ singularity is not included in the space and, of course, in this system it is obvious the metric is not singular at $r=4m$. For this argument, see \cite{gib:77}.}. 
\begin{figure}
\begin{center}
\includegraphics[height=1in, width=6in]{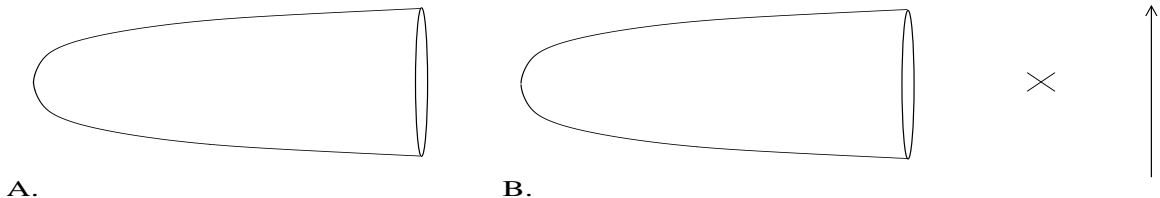}
\caption{The bolt solution. A: The Euclidean black hole, the periodic direction depicts the Euclidean time. The $\theta$ and $\phi$ direction have been suppressed. B: The Kaluza bolt soliton, the periodic direction is now the direction of compactification $x^5$. The Kaluza bolt can be interpreted as a Euclidean black hole multiplied with a trivial time coordinate.}
\end{center}
\end{figure}
Again we can incorporate this gravitational instanton as a soliton in the $n=0$ Kaluza-Klein theory, in the same manner as before, namely by just adding trivially the time component \cite{gro:83, cho:82}:
\be 
ds^2 = -dt^2 + (1-4m/r) (dx^5)^2 + (1-4m/r)^{-1} dr^2 + r^2 d\Omega^2. 
\ee 
Just as for the monopole, the regularity condition (at $r=4m$) imposes the space to have a compact fifth dimension with $x^5$ having a period of $16\pi m$. At $r=4m$ the spacetime is pinched off as the radius of the fifth dimension shrinks to zero (the set of fixed points of $\p/ \p x^5$ at $r=4m$ defines what is called a 'bolt'). As $r \rightarrow \infty$  the radius of the fifth direction remains of finite size (see also figure 2). 

One feature of Kaluza solitons we would like to mention in particular, namely that all $n=0$ Kaluza-Klein solitons written down by Gross and Perry \cite{gro:83} exhibit peculiar gravitational properties. For instance, the bolt has an inertial mass, given by:
\be 
P^{0}=\int d^3 x dx^5 \sqrt{-g^{5}}\, R^{00} =\frac{8 \pi}{\kappa}m.
\ee
Both examples of solitons we have given are however 'flat' in the time direction, so in these spaces there are time-like geodesics that correspond to a particle sitting at rest relative to the soliton ($dx^{A}/d\tau=k\delta^{A0}$). The Newtonian force ($F \propto \frac{1}{2}\nabla g_{00}$) a test particle experiences, vanishes for these spaces. Hence the solitons have a zero gravitational mass:
\be 
M_{grav}=0.
\ee
An observer at a fixed point in space cannot distinguish whether a soliton is present or not. This is argued not to be a violation of the (five dimensional) principle of equivalence, but of the (four dimensional) Birkhoff theorem - in the sense that not every rotation symmetric $5\times 5$ metric has at large distances the structure of the Schwarzschild solution in its $g^{(4)}_{\mu \nu}$ part \cite{gro:83}. The absence of gravitational mass actually makes it possible for these spaces to be non-singular: the Hawking-Penrose singularity theorems can be extended to five dimensions \cite{bee:96}, but they do not apply as no trapped surface, in the interior of some horizon, is formed.

\subsection{Scale invariance}

Now why would the Einstein-Pauli argument not seem to apply to these soliton spaces? Equivalently, one may wonder how the gravitational instantons stand the comparison with Einstein's four dimensional argument \cite{ein:42}. Note that Einstein's Assumption A allows a Euclidean signature and a periodic time dependence. But Einstein tacitly assumed the period and the mass parameter to be independent. In the case of the gravitational instantons considered, however, the period is related to the mass parameter. 
Therefore, Einstein's contradiction in terms does not catch on, since precisely replacing $m$ by $(1+c)m$ would correspond to performing a dilation on the Euclidean time $\tau$. In the case of the Euclidean black hole the dilation produces a conical singularity with an angular deficit of $16\pi cm$ at the horizon. So the dilation is not a symmetry of these solutions. In other words: the Euclidean black hole solution has action $S=16 \pi m^2$ \cite{gib:79} and consequently under the dilation $m \rightarrow (1+c) m$, the action is not stationary.

For basically the same reason the Einstein and Pauli argument falls short for the five dimensional solitons described here. Notably, the Kaluza bolt solution (\ref{eq:bolt}) has the right asymptotics to fit their argument (\ref{eq:sy}):
\be
ds^2 \sim -dt^2 + \Big( 1-\frac{4m}{r} \Big) (dx^5)^2 + \Big(1+\frac{4m}{r}\Big)(dx^i)^2 \,\,\,\,\,\,\,\,\, \mbox{as} \,\,\,\,\, r \rightarrow \infty
\ee
which complies (\ref{eq:sy},\ref{eq:demd}) with $m_{00}=m_{05}=0$ and $m_{55}=-m$. $m_{55}$ had been shown to be necessarily zero by Einstein and Pauli on basis of the equation;
\be
c^{5}_{5} \oint \sqrt{|g^{5}|}g^{5 A} \Gamma^{i}_{5 A} n_{i}d^2 \Omega=0
\ee 
following from the assumed symmetry transformation
\be 
x^5 \rightarrow x^5 + \xi^5 = x^5 + c^{5}_{5} x^5     \label{eq:eindel}
\ee
i.e. a dilation in the fifth dimension. If the bolt is to be considered non-singular one must have a compact fifth direction: $x^5$ is taken periodic with period $16 \pi m$. 
As a consequence one has lost scale invariance: dilations in $x^5$ are no longer symmetry transformations because the invariant radius $R$ changes under these transformations:
\be 
R = \int dx^5 \sqrt{g_{55}}=2\pi m \sqrt{g_{55}(r)}  \rightarrow (1+c) R.
\label{eq:invarvar}
\ee
A particular solution can have a period $m$ or a period $(1+c)m$ at infinity, but if the periods differ they label distinct solutions that are not equivalent by virtue of global scale symmetry. In the bolt solution, dilations on $x^5$ would introduce conical singularities. 
These transformations therefore cannot be considered a symmetry of this soliton: one would allow the scale invariance and the associated conservation law Einstein and Pauli use, the fifth direction cannot be compact and the bolt metric can only be considered singular. The Kaluza bolt has no curvature in the time direction, so its action is proportional to the instanton action: the action of the five dimensional object is thus again not stationary in dilations in the compact direction.

The same statement can be made in case of the nut (i.e. the Kaluza-Klein monopole - note that it does not have the asymptotics (\ref{eq:asympt})), with action (in the four space-like dimensions) $16\pi m^2$ \cite{gib:79}. Also here, when giving up the compactification of the fifth direction to allow dilation invariance, one would have to face the Dirac string singularity again. The compactification is required in order for both spaces to be described regular, but this violates the dilation invariance on which Einstein and Pauli's argument is based. Time dilation is still a symmetry in five dimensions, as this is not a compact direction. It is the conservation law associated with this symmetry that made Einstein and Pauli conclude $m_{00}=0$. This can explain why there are no solitons in Kaluza's theory (for which one requires general covariance in the four space-time dimensions) that have a gravitational mass (not quite surprising as this mass would introduce a horizon).

One could look at the theorem given by Einstein and Pauli as in fact a positive, rather than a negative result, as considered by Einstein. Namely, it would show that there are no non-singular particle solutions if one retains the vacuum Kaluza-theory and that in order to find these objects one needs a more general theory, a theory like classical Kaluza-Klein as described by Einstein and Bergmann. Nevertheless, we expect Einstein would rather be discouraged by the apparent impossibility of solitons in Kaluza's theory. He never worked on a five dimensional theory again, and it is quite clear that also in the full Kaluza-Klein theory Einstein's search for a non-singular particle had become frustrated. 

\section{Conclusion}

We have seen that central in Einstein's work on either Kaluza theory or the full classical Kaluza-Klein was his desire to describe an electro-magnetically interacting object that would be a non-singular solution of the field equations. In his papers he expressed the desire to find such an object, just as he mentions this search was repeatedly frustrated. When in Kaluza's theory (with constant dilaton) he does not find a non-singular bridge type solution, he starts working on the expanded Kaluza-Klein theory (with constant dilaton) \cite{ein:38}. Here again we learned, in particular by studying his correspondence with his assistants, that he foremost wanted to find a non-singular electrically charged particle. He did not succeed \cite{ein:41}, which is probably one of the primary reasons why he abandoned working on this theory. In 1943 he argued, together with Pauli, that in Kaluza's theory (now with variable dilaton) it would principally be impossible to find a non-singular particle. Einstein never worked in five dimensions again.

Striking is that solitons actually exist in this theory. The description of electro-magnetically interacting solitons requires of course a quite modern point of view (i.e. for the magnetic monopole Kaluza's theory as a non-trivial principal bundle). Taking that position one can argue Einstein, in his discussions with Bergmann and in his paper with Pauli, had too restricted boundary conditions. One can only guess whether these objects are the exact sort of particles Einstein was looking for, however much in close contact the Einstein-Pauli argument stands to these modern day solitons. Indeed they are regular, massive non-singular particle objects that solve the classical field equations and they show Einstein's intuition was not far off. With these solitons one can realize classical particles in a non-singular way - and there is no quantum state in the description of the particle involved. Anyway, we feel that Einstein's search for such solitonic objects should not be isolated from his attitude regarding quantization and his possible plans to re-derive the quantum relations.  

Indeed, we find the most important lesson one learns about 
Einstein, when working on Kaluza-Klein, is that he would never even mention the result Klein had arrived at: explaining the discrete charge spectrum by virtue of the de Broglie relation together with the compactification of the fifth direction. Not once in his considerations, for example concerning the scale of the fifth direction, would he use the de Broglie relation that is so central to Klein's reasoning. He instead might have wanted to reason the other way around: after having found the desired solitonic objects, that by virtue of the boundary conditions would give a discrete charge spectrum, it would be possible to show that there had to be a minimum of action $h$. In any case, by his continual ignoring of Klein's argument one sees that when unifying the existence of discretised matter with the ideal of continuous fields, Einstein would want the quantum relations to follow, not to guide. 

This attitude regarding quantization did not change after Einstein had left all the varieties of the Kaluza-Klein theory. His assistant Bergmann, however, would change position concerning quantization. He would later work on quantum gravity, on a theory of gravity in which even the gravitational interaction would be quantized in advance. In their later correspondence, Einstein exhibits again his discomfort with the ad hoc quantization procedures. In 1949 (just when Einstein was recovering from a laparotomy) Bergmann, now in Syracuse, wrote Einstein and asked if they could have a discussion sometime:\\
\\
\emph{``As anyone can only be a crank about his own ideas, and as you are someone who combines steadfastness with the ability to acknowledge his hypothesis could go wrong (usually one can only find just one of these qualities, mostly the latter) I would appreciate very much talking to you and hearing your observations; whether we appreciate the same or not, what we want is sufficiently related that we could easily come to an understanding.''} \cite{ber:49} \footnotemark[2]\\
\\
Einstein replies:\\
\emph{``You are looking for an independent and new way to solve the fundamental problems. With this endeavor no one can help you, least of all someone who has somewhat fixed ideas. For instance, you know that on the basis of certain considerations I am convinced that the probability concept should not be primarily included in the description of reality, whereas you seem to believe that one should first formulate a field theory and subsequently 'quantize' it. This is in keeping with the view of most contemporaries. Your effort to abstractly carry through a field theory without having at your disposal the formal nature of the field quantities in advance, does not seem favorable to me, for it is formally too poor and vague.''}\footnote{Einstein continues with: \emph{``Jedenfalls kann ich Sie also f\"ur die absehbare Zukunft nicht zu einem Besuch einladen und versichere Ihnen nochmals, dass einem bei der Eierlegerei niemand helfen kann, sondern das Gesch\"opf in Gottesnamen allein dasteht.''} \cite{ein:91}} \cite{ein:91} \footnotemark[2]\\
\\
In 1954 Bergmann asks for funding at the National Science Foundation. Einstein is asked to evaluate the request, entitled ``quantum theory of gravitation'':\\
\emph{``The application of Dr. Bergmann concerns a problem of central significance for modern physics. All physicists are convinced of the high truth value of the probabilistic quantum theory and of the general relativity theory. These two theories, however, are based on independent conceptual foundations, and their combination to a unified logical system has so far resisted all attempts in this direction...If the decision were mine I should grant the funds asked for by Dr. Bergmann, in view of the central importance of the problem and the qualifications of the candidate. Even though in my opinion the probability of attaining the great goal seems rather small at this point, the financial risk incurred is on the other hand so modest that I should have no qualms to grant the application.''} \cite{ein:90}\\
\\
Finally, in 1955 Einstein writes:\\
\emph{``Is it conceivable that a field theory permits one to understand the atomistic and quantum structure of reality? Almost everybody will answer this question with 'no.' But I believe that at the present time nobody knows anything reliable about it. This is so because we cannot judge in what manner and how strongly the exclusion of singularities reduces the manifold of solutions. We do not possess any method at all to derive systematically solutions that are free of singularities...we cannot at present compare the content of a nonlinear field theory with experience...I see in this method} [i.e. quantization] \emph{only an attempt to describe relationships of an essential nonlinear character by linear methods.''} \cite{ein:55}   \\
\\
\\
\bfseries{Acknowledgments} \mdseries

I profited greatly from discussions with Herman Verlinde, Michel Janssen, Anne Kox and Cristoph Lehner, and very much enjoyed the hospitality of the Einstein Papers Project at Boston University in the autumn of 1999. I am also greatful for remarks by David Gross. Finally I thank the Hebrew University of Jerusalem for permission to publish.

\small
\bibliography{einkalklein}

\begin{thebibliography}{10}

\bibitem{bai:83}
F.A. Bais.
\newblock Dirac monopoles, from d=2 to d=5.
\newblock In {\em Progress in Gauge field theory}. Plenum, 1983.

\bibitem{bee:96}
J.~Beem, P.~Ehrlich, and K.~Easley.
\newblock {\em Global {L}orentzian {G}eometry}.
\newblock Marcel Dekker, 1996.

\bibitem{ber:38}
P.~Bergmann.
\newblock Letter to {A.} {Einstein}, 15 {August} 1938, {AE} 6-272.

\bibitem{ber:49}
P.~Bergmann.
\newblock Letter to {A.} {Einstein}, 24 {January} 1949, {AE} 6-282.

\bibitem{ber:37}
P.~Bergmann.
\newblock Letter to {A.} {Einstein}, 4 {August} 1938, {AE} 6-270.

\bibitem{ber:48}
P.~Bergmann.
\newblock Unified field theory with fifteen field variables.
\newblock {\em Ann. Math.}, 49:255, 1948.

\bibitem{cho:82}
A.~Chodos and S.~Detweiler.
\newblock Spherically symmetric solutions in five-dimensional relativity.
\newblock {\em Gen. Rel. Grav.}, 14:879, 1982.

\bibitem{dir:31}
P.~Dirac.
\newblock Quantised singularities in the electromagnetic field.
\newblock {\em Proc. Roy. Soc. London A}, 133:60, 1931.

\bibitem{duf:86}
M.~Duff, B.~Nilsson, and C.~Pope.
\newblock {Kaluza-Klein} {supergravity}.
\newblock {\em Phys. Rep.}, 130:1, 1986.

\bibitem{ear:99}
J.~Earman and J.~Eisenstaedt.
\newblock Einstein and singularities.
\newblock {\em Stud. Hist. Phil. Mod. Phys.}, 30:185, 1999.

\bibitem{ein:94}
A.~Einstein.
\newblock Letter to {H.} {Weyl}, 6 {June} 1922, {AE} 24-71.

\bibitem{ein:90}
A.~Einstein.
\newblock Letter to {National} {Science} {Foundation}, 18 {April} 1954, {AE}
  6-313.

\bibitem{ein:96}
A.~Einstein.
\newblock Letter to {P.} {Bergmann}, 21 {June} 1938, {AE} 6-249.

\bibitem{ein:91}
A.~Einstein.
\newblock Letter to {P.} {Bergmann}, 26 {January} 1949, {AE} 6-283.

\bibitem{ein:97}
A.~Einstein.
\newblock Letter to {P.} {Bergmann}, 4 {July} 1938, {AE} 6-253.

\bibitem{ein:98}
A.~Einstein.
\newblock Letter to {P.} {Bergmann}, 5 {August} 1938, {AE} 6-271.

\bibitem{ein:95}
A.~Einstein.
\newblock Letter to {V.} {Bargmann}, 9 {July} 1939, {AE} 6-207.

\bibitem{ein:15}
A.~Einstein.
\newblock Die {Grundlage} der allgemeinen {Relativit{\"a}tstheorie}.
\newblock {\em Ann. Phys.}, 49:769, 1916.

\bibitem{ein:16}
A.~Einstein.
\newblock {N{\"a}herungsweise} {Integration} der {Feldgleichungen} der
  {Gravitation}.
\newblock {\em Sitzungsber. Preuss. Akad. Wiss.}, Phys. Math. Kl.:688, 1916.

\bibitem{ein:20b}
A.~Einstein.
\newblock {\em {\"A}ther und {R}elativit{\"a}tstheorie}.
\newblock Springer, 1920.

\bibitem{ein:20}
A.~Einstein.
\newblock Remark in: {Votr{\"a}ge} und {Diskussionen} von der 86.
  {Naturvorscherversammlung} in {Nauheim} vom 19.-25. {September} 1920.
\newblock In {\em Phys. Zs.}, volume~21, page 650, 1920.

\bibitem{ein:21}
A.~Einstein.
\newblock Geometrie und {Erfahrung}.
\newblock {\em Sitzungsber. Preuss. Akad. Wiss.}, Phys. Math. Kl.:123, 1921.

\bibitem{ein:23}
A.~Einstein.
\newblock Bietet die {Feldtheorie} {M{\"o}glichkeiten} f{\"u}r die {L{\"o}sung}
  des {Quantenproblems}?
\newblock {\em Sitzungsber. Preuss. Akad. Wiss.}, Phys. Math. Kl.:359, 1923.

\bibitem{ein:27}
A.~Einstein.
\newblock Zu {K}aluza's {Theorie} des {Zusammenhanges} von {Gravitation} und
  {E}lektrizit{\"a}t (erste und zweite {M}itteilung).
\newblock {\em Sitzungsber. Preuss. Akad. Wiss.}, Phys. Math. Kl.:23, 1927.

\bibitem{ein:42}
A.~Einstein.
\newblock Demonstration of the non-existence of gravitational fields with a
  non-vanishing total mass free of singularities.
\newblock {\em Revista Universidad Nacional de Tucuman}, A2:11, 1941.

\bibitem{ein:55}
A.~Einstein.
\newblock {\em The meaning of relativity}.
\newblock Princeton UP, 1955.
\newblock Fifth enlarged edition.

\bibitem{ein:99}
A.~Einstein.
\newblock Letter to {W.} {Pauli}, 19 {September} 1938, {AE} 19-175.
\newblock In {\em Wolfgang {Pauli} - {Wissenschaftlicher} {Briefwechsel},
  {Band} {II}, 1930-1939}. Springer, 1985.

\bibitem{ein:41}
A.~Einstein, V.~Bargmann, and P.~Bergmann.
\newblock On the five-dimensional representation of gravitation and
  electricity.
\newblock In {\em Theodore von Karman Anniversary Volume}. California Institute
  of Technology, 1941.

\bibitem{ein:38}
A.~Einstein and P.~Bergmann.
\newblock On a generalisation of {Kaluza's} theory of electricity.
\newblock {\em Ann. Math.}, 34:683, 1938.

\bibitem{ein:22}
A.~Einstein and J.~Grommer.
\newblock Beweis der {Nichtexistenz} eines {\"u}berall regul{\"a}ren zentrisch
  symmetrischen {Feldes} nach der {Feld-theorie} von {Th.} {Kaluza}.
\newblock {\em Scripta Universitatis atque Bibliothecae Hierosolymitanarum:
  Mathematica et Physica}, 1:1, 1923.

\bibitem{ein:43}
A.~Einstein and W.~Pauli.
\newblock On the non-existence of regular stationary solutions of relativistic
  field equations.
\newblock {\em Ann. Math.}, 44:131, 1943.

\bibitem{ein:35}
A.~Einstein and N.~Rosen.
\newblock The particle problem in the general theory of relativity.
\newblock {\em Phys. Rev.}, 48:73--77, 1935.

\bibitem{gib:77}
G.~Gibbons and S.~Hawking.
\newblock Action integrals and partition functions in quantum gravity.
\newblock {\em Phys. Rev.}, D15:2752, 1977.

\bibitem{gib:79}
G.~Gibbons and S.~Hawking.
\newblock Classification of gravitational instanton symmetries.
\newblock {\em Comm. Math. Phys.}, 66:292, 1979.

\bibitem{gro:83}
D.~Gross and M.~Perry.
\newblock Magnetic monopoles in {Kaluza-Klein} theories.
\newblock {\em Nucl. Phys.}, B226:29, 1983.

\bibitem{haw:77}
S.~Hawking.
\newblock Gravitational instantons.
\newblock {\em Phys. Lett.}, 60A:81, 1977.

\bibitem{haw:78}
S.~Hawking.
\newblock Euclidean {Quantum} {Gravity}.
\newblock In {\em Recent Developments in Gravity}. Plenum, 1978.

\bibitem{jor:47}
P.~Jordan.
\newblock Erweiterung der projektiven {Relativit\"atstheorie}.
\newblock {\em Ann. Phys.}, 1:219, 1947.

\bibitem{kal:21}
Th. Kaluza.
\newblock Zum {Unit{\"a}tsproblem} der {Physik}.
\newblock {\em Sitzungsber. Preuss. Akad. Wiss.}, Phys. Math. Kl.:966, 1921.

\bibitem{kle:27}
O.~Klein.
\newblock The atomicity of electricity as a quantum theory law.
\newblock {\em Nature}, 118:516, 1926.

\bibitem{kle:26}
O.~Klein.
\newblock Quantentheorie und f{\"u}nfdimensionale {Relativit{\"a}tstheorie}.
\newblock {\em Z. Phys.}, 37:895, 1926.

\bibitem{lic:38}
A.~Lichnerowicz.
\newblock Sur certaines probl{\`e}mes globaux relatifs au syst{\`e}me des
  {\'e}quations d'{E}instein.
\newblock Thesis, Paris 1939.

\bibitem{lic:39}
A.~Lichnerowicz.
\newblock Espaces temps ext{\'e}rieurs r{\'e}guliers partout.
\newblock {\em C. R. Acad. Sci. Paris}, 206:313, 1939.

\bibitem{lic:46}
A.~Lichnerowicz.
\newblock Sur le charact{\`e}re euclidien d'espaces-temps ext{\'e}rieurs
  statiques partout r{\'e}guliers.
\newblock {\em C. R. Acad. Sci. Paris}, 222:432, 1946.

\bibitem{lic:55}
A.~Lichnerowicz.
\newblock {\em Th{\'e}ories relativistes de la gravitation et de
  l'{\'e}lectromagn{\'e}tisme}.
\newblock Masson, 1955.

\bibitem{lic:45}
A.~Lichnerowicz.
\newblock Letter to {W.} {Pauli}, 6 oktober 1945.
\newblock In {\em Wolfgang {Pauli} - {Wissenschaftlicher} {Briefwechsel},
  {Band} {III}, 1940-1949}. Springer, 1985.

\bibitem{mis:63}
C.~Misner.
\newblock The flatter regions of {Newman}, {Unti} and {Tamburino's} generalised
  {Schwarzschild} space.
\newblock {\em J. Math. Phys.}, 4:924, 1963.

\bibitem{str:98}
L.~O'Raifeartaigh and N.~Straumann.
\newblock Gauge theory: historical origins and some modern developments.
\newblock {\em Rev. Mod. Phys.}, 72:1, 2000.
\newblock hep-ph/9810524.

\bibitem{pau:20}
W.~Pauli.
\newblock Remark in: {Vortr{\"a}ge} und {Diskussionen} von der 86.
  {Naturvorscherversammlung} in {Nauheim} vom 19.-25. {September} 1920.
\newblock In {\em Phys. Zs.}, volume~21, page 650, 1920.

\bibitem{pau:33}
W.~Pauli.
\newblock {\"U}ber die {F}ormulierung der {N}aturgesetze mit f{\"u}nf homogenen
  {Koordinaten. {Teil} {I}. (klassischen {Theorie}) }.
\newblock {\em Ann. Phys.}, 18:305, 1933.

\bibitem{pau:34}
W.~Pauli.
\newblock {\"U}ber die {F}ormulierung der {N}aturgesetze mit f{\"u}nf homogenen
  {Koordinaten.} {Teil} {II}. ({Die} {Diracschen} {Gleichungen} {f\"ur} die
  {Materiewellen}).
\newblock {\em Ann. Phys.}, 18:305, 1933.

\bibitem{pau:58}
W.~Pauli.
\newblock {\em Theory of relativity}.
\newblock Pergamon, 1958.
\newblock Translation of German original of 1921.

\bibitem{pau:38}
W.~Pauli.
\newblock Letter to {A.} {Einstein}, {September} 6, 1938, {AE} 19-174.
\newblock In {\em Wolfgang {Pauli} - {Wissenschaftlicher} {Briefwechsel},
  {Band} {II}, 1930-1939}. Springer, 1985.

\bibitem{des:83}
V.~De Sabatta and E.~Schutzer (eds.).
\newblock {\em Unified field theories of more than four dimensions}.
\newblock World Scientific, 1983.

\bibitem{sor:83}
R.~Sorkin.
\newblock Kaluza-{K}lein monopole.
\newblock {\em Phys. Rev. Lett.}, 51:87, 1983.

\bibitem{thi:48}
Y.~Thiry.
\newblock Les {\'e}quations de la th{\'e}orie unitaire de {Kaluza}.
\newblock {\em C. R. Acad. Sci. Paris}, 226:216, 1948.

\bibitem{tip:80}
F.~Tipler, C.~Clarke, and G.~Ellis.
\newblock Singularities and horizons - a review article.
\newblock In {\em General relativity and gravitation, vol 2}. Plenum, 1980.

\bibitem{viz:94}
V.~Vizgin.
\newblock {\em Unified field theories in the first third of the 20th century}.
\newblock Birkh{\"a}user, 1994.

\bibitem{wuy:75}
T.~Wu and C.~Yang.
\newblock Concept of nonintegrable phase factors and global formulation of
  gauge fields.
\newblock {\em Phys. Rev.}, D12:3845, 1975.

\end{thebibliography}
\bibliographystyle{plain}
\include{einkalklein.bib}

\end{document}